\documentclass[intlimits,twoside,a4paper]{article}

\usepackage{amsmath,amssymb}
\usepackage{graphicx}

\usepackage[T2A]{fontenc}
\usepackage[cp1251]{inputenc}

\usepackage[eqsecnum]{cmpj2}

\issue{2016}{19}{1}{13605}
\doinumber{10.5488/CMP.19.13605}

\title[Equilibrium clusters]%
{Equilibrium clusters in suspensions of
colloids interacting via potentials with a
local minimum
}

\author[A. Baumketner, W. Cai]{A. Baumketner\refaddr{label1}\thanks{E-mail:  andrij@icmp.lviv.ua}\,,
        W. Cai\refaddr{label2}}
\addresses{
\addr{label1} Institute for Condensed Matter Physics, National Academy of Sciences of Ukraine, \\ 1~Svientsistskii St.,  79011 Lviv, Ukraine
\addr{label2} Beijing Computational Science Research Center, Beijing 100094, China
}

\authorcopyright{A. Baumketner, W. Cai, 2016}
\date{Received  November 10, 2015, in final form December 18, 2015}

\newcommand{\ds}{\displaystyle}
\begin{document}

\maketitle

\begin{abstract}
In simple colloidal suspensions, clusters  are various multimers that result from
colloid self-association and exist in equilibrium with monomers.There are two
types of potentials that are known to produce clusters: a) potentials
that result from the competition
between short-range attraction and long-range repulsion
and are characterized by a
\textit{global} minimum and a repulsive tail
and b) purely repulsive
potentials which have a soft shoulder. Using computer simulations, we demonstrate
in this work that potentials with a \textit{local} minimum and a repulsive tail,
not belonging to either of the known types, are also capable of generating clusters. A
detailed comparative analysis shows that the new type of cluster-forming
potential serves  as a bridge between the other two. The new clusters are
expanded in shape and their assembly is driven by entropy, like in  the purely
repulsive systems but only at low density. At high density, clusters are
collapsed and stabilized by energy, in common with the systems with competing
attractive and repulsive interactions.

\keywords colloids, clusters, local minimum, repulsive potential, computer simulations
\pacs 64.75.Yz, 61.20.Gy, 61.20.Ja, 61.20.Ne, 05.10.-a

\end{abstract}

\section{Introduction}

The term ``clusters''
 refers to a large variety of objects that range in size from small multimers to
mesoscopic domains~\cite{r1} and arise as a consequence of monomer self-association in a large variety of soft
materials~\cite{r2}. Most often, clusters are
discussed in reference to colloidal suspensions~\cite{r3}, where they exist
in equilibrium with monomers, but they were also reported for proteins~\cite{r4},
synthetic clays~\cite{r5} and metal
nanoparticles~\cite{r6}.
Equilibrium clusters become the dominant species in the solution at appropriate
thermodynamic conditions. They may also arise transiently, as a consequence of arrested phase
transition~\cite{r7}.

As a particular case of the self-assembly process, cluster formation is of
key interest to basic research, in
particular condensed matter physics. Additionally, it also has appreciable practical applications, for
instance as a drug-delivery vehicle~\cite{r8}.
Clusters are capable of significantly altering the mechanical properties of
aqueous solutions in which they assemble. This is the case, for instance, of solutions containing
monoclonal antibodies, a known biopharmaceutical, which experience a considerable viscosity increase
if clusters are present~\cite{r9}. It is
imperative, therefore, to develop a basic understanding of the principles
underlying the formation of clusters, in order to use these systems successfully for therapeutic
purposes.

Historically, it seems, the possibility of clusters emerging spontaneously in a homogeneous fluid was first
raised within the concept of competing interactions. When discussing phase transitions in systems
interacting via attractive potentials, Lebowitz and Penrose~\cite{r10}
 asked about the effect of an additional
repulsive part in the potential, whose range is longer than that of the attractive part. Their conclusion
was that the liquid formed as a result of the normal first-order gas-liquid transition due to the attraction
among particles would break into finite-size droplets whose size is large compared to the size of the
colloid but small compared to the range of the repulsion. Over the past forty years, this scenario was
seen to play out in a variety of systems, interacting by a variety of potentials. Kendrick~ et al.~\cite{r11}
considered a system of colloids interacting via repulsive Coulomb and attractive van~der~Waals forces
and concluded that as a result of the competition between these two interactions, the critical point
between the low density liquid phase and the high-density liquid phase is `preempted by a finite wave
vector critical point'~\cite{r11}. As a result of the new phase transition, macrophases with inhomogeneous
density distributions are formed. The same idea of competition between short-range attractive and
long-range repulsive forces (SALR) in simple fluids was later researched by Sear~et~al.~\cite{r12,r13}
 in the
context of colloids lying on the air-water interface. Analytic mean-field theory~\cite{r13} and computer
simulations~\cite{r12}
indicated that, indeed, instead of the liquid-liquid phase separation (LLPS), colloids form
various patterned (modulated) phases, including finite-size clusters. In many respects, these were
similar to the patterns created by the competition between repulsion and attraction and described
previously for complex fluids~\cite{r14,r15}. Further progress was made with the work of Groenewold and
Kegel~\cite{r16} who showed how an elaborated model describing the amount of charge on a colloidal surface
may lead to the stabilization of clusters of large size. Since that publication, a large amount of work on
SALR clusters has followed (see references~\cite{r17,r18} and references therein) focusing mainly on how the LLPS
line is broken/modified by the presence of the repulsion in the potential. A sophisticated and accurate
self-consistent Ornstein-Zernike approximation (SCOZA) within the integral-equation theory of the liquid
state was used by the group of Reatto~\cite{r19,r20} to delineate the line in the phase diagram separating the
homogeneous fluid phase from the cluster fluid phase, or the so-called $\lambda$-line. The SCOZA predictions were
later compared to the results of a density-functional theory~\cite{r21} and simulation~\cite{r22}.
A partial list of other
topics that have been covered includes: a) the effect of the attraction range~\cite{r23,r24},
b) the height of the
barrier~\cite{r25}, and c) the role of the tail in the potential~\cite{r26,r27,r28,r29,r30}.

The second class of systems for which clusters have been reported are colloids interacting via repulsive
potentials with a soft shoulder~\cite{r2}. Stell and Hemmer pointed out in a seminal
work~\cite{r31} that due to
the additional length scale, one set by the size of the colloid and the other by the size of the shoulder,
such potentials possess very unusual properties. In addition to the usual gas-liquid transition ending in a
critical point, for instance, potentials with shoulders may exhibit a liquid-liquid transition, which also
ends in a critical point. There have been numerous follow-up studies that focused on the specific details
of the new phase transitions as well as on whether or not
this model is capable of explaining the thermodynamic
anomalies of liquid water~\cite{r32,r33,r34,r35,r36}. It was not until the work of Klein~et~al.~\cite{r37}
 that clusters were made
the specific subject of such studies. Using analytical theory and computer simulations, colloids
interacting via a hard-core and soft shoulder (HCSS) potentials were seen in this work to form spherical
clusters (or clumps) in the fluid phase. At low temperature, clusters were observed to freeze into cluster
crystals, cluster glasses and a number of polycrystalline materials~\cite{r37}.
In a later study for the same
potential but in 2D, Norizoe and Kawakatsu~\cite{r38,r39} also reported clusters that are string-like or
extended in shape. Extended, chain-like clusters were also observed by Camp~\cite{r40} for a continuous
repulsive potential with a shoulder (different from that of HCSS); the shoulder was determined to be
responsible for the emergence of the clusters. In addition to clusters, HCSS are capable of forming a variety of
other ordered phases at low temperature. Malescio and Pellicane~\cite{r41} reported a regular striped phase
while Glaser~et~al.~\cite{r42} discovered cluster crystals and cluster fluids in MC simulations.
 Cluster fluids were
also reported by Mladek~et~al.~\cite{r43}, lanes by Fornleitner and Kahl~\cite{r44},
and lamella, hexagonal-columnar
and body-center cubic phases, as well as the associated inverse structures by Shin~et~al.~\cite{r45}.
Dotera~et~al.~\cite{r46}
describe various quasicrystalline structures that are formed in 2D at low temperature.

It should be noted that many of the properties displayed by the cluster-forming colloids with repulsive
interactions actually do not require the presence of the hard core. Likos~et~al.~\cite{r47} considered
a number
of potentials bound at the origin, including the flat portion of the HCSS lacking the hard core, and
introduced a criterion for determining if a system with such potential should experience a micro-phase
separation into a lattice state where multiple colloids occupy the same lattice site. This criterion was
later successfully tested in computer simulations~\cite{r48}, which indeed uncovered cluster crystals.
Repulsive~\cite{r3,r7} as well as attractive~\cite{r49} colloids
were seen to form clusters in laboratory experiments.

In this work we demonstrate by computer simulations that potentials that do not belong to either of
the classes mentioned above are also capable of forming equilibrium clusters. Specifically, we consider
the potentials that, instead of the global minimum as in the SALR scheme, display a local minimum that has a
positive energy and is separated from the longer-distance, zero-energy states by a finite barrier. Unlike
in the SALR scheme, however, potentials with local attractive minima and long-range repulsion (LALR)
lack the energy incentive for the particles to self-associate. LALR model shares with the HCSS potential
the global repulsive character and soft shoulder. Experimentally, such potentials may arise as a result of
incomplete cancellation between the attractive and repulsive terms. Theoretically, potentials with local
minima are little studied. Batten~et~al.~\cite{r50} showed that such potentials have unique ground-state
configurations, including kagome and honeycomb crystals, and stripes. Liu~et~al.~\cite{r51}
 also examined
ground-state configurations of a LALR system confined to a plane. Neither of these papers focused on
clusters. To the best of our knowledge, this subject has never been studied.

A comparative analysis of clusters formed by systems of all three schemes, LALR, SALR and HCSS,
indicates that the LALR serves as a bridge between the other two. Like in the HCCS model, LALR
potentials induce entropy-driven expanded clusters but only at low overall system density and for small
cluster sizes. With an increase in the density, the LALR system, like its SALR counterpart, leads to
energy driven clusters of large size. The dual identity of the LALR model is also revealed in its
temperature behavior. Small clusters of this system are stabilized by temperature, like in the HCSS, but
larger clusters are destabilized, in common with the SALR.

\section{Methods}
\begin{figure}[!t]
\centering
\includegraphics[width=0.6\textwidth]{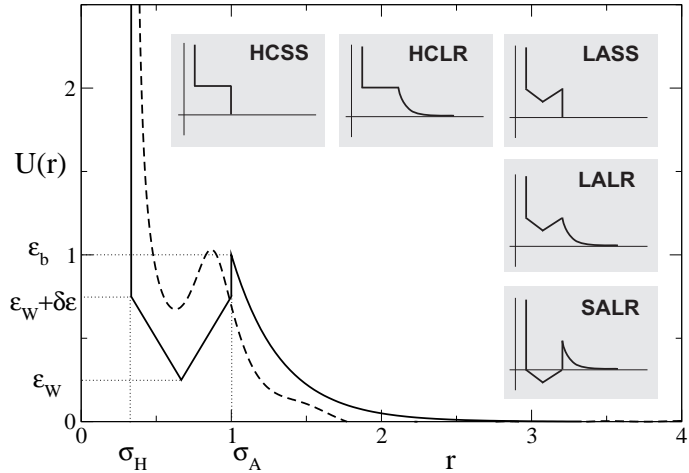}
\caption{
Schematic representation of the potential considered in this work. Depending on the choice of
parameters, different types of soft-matter systems can be studied. Five specific shapes investigated in detail
are shown as icons. Broken line shows potential, appropriately scaled, obtained for
lysozyme solutions at acidic pH.
}
\label{f7}
\end{figure}

The shape of the model potential considered in this work is inspired by the inter-molecular potential
of aqueous solutions of lysozyme. Using structural functions available from
scattering experiments for this protein, effective
potentials were derived in our prior work (data not published) for
a number of pH values. For pH 2, the potential was seen to
possess a local minimum at short distances followed by a repulsive
term decaying to zero at longer distances. To reflect these features, the following analytical
function was considered:
\begin{equation}
u(r)=\left\{
\begin{array}{lll}
+\infty, & r<\sigma_\text{H}, &  \\
-a r + b + \epsilon_w, & \sigma_\text{H} < r < \ds {\frac{\sigma_\text{H} + \sigma_A}{2}}, &
a=\ds {\frac{2 \delta \epsilon}{\sigma_A - \sigma_\text{H} }}, \\[2ex]
a r - b + \epsilon_w, & \ds {\frac{\sigma_\text{H} + \sigma_A}{2}} < r < \sigma_A, &
b=\delta \epsilon \ds {\frac{\sigma_A + \sigma_\text{H} }{\sigma_A - \sigma_\text{H} }}, \\
\epsilon_b \sigma_A  \ds { \frac{\re^{-\kappa ( r -\sigma_A ) }}{r}}, & \sigma_A < r. &
\end{array}
\right.
\end{equation}
The shape of the potential is determined by a set of six basic parameters,
$\sigma_\text{H}$, $\sigma_A$, $\epsilon_w$, $\epsilon_b$, $\delta \epsilon$ and $\kappa$.
A hard-core wall is placed at $\sigma_\text{H}$. The short-range (SA) part
of the potential contains a minimum
(either local or global) located between $\sigma_\text{H}$ and $\sigma_A$. The
minimum is triangular in shape. Its energy is
defined by $\epsilon_w$  while the depth is given by $\delta \epsilon$.
The long-range part (LR), defined by distances greater
than $\sigma_A$ is either missing or given by the Yukawa potential characterized
by the inverse decay length $\kappa$.
The SA and LR parts meet at $r = \sigma_A$ where the energy is $\epsilon_b$.
The schematic of the model potential is shown
in figure~\ref{f7} together with its experimental prototype.
Depending on the choice of parameters, the
potential may represent a number of soft-matter systems. Throughout this article, we
use $\sigma_A$ as the unit of distance and $\epsilon_b$ as the unit of energy. The
hard-core diameter is set at $\sigma_\text{H} = \sigma_A /3$. The remaining three
parameters are varied to achieve the following five representative
shapes: 1) Hard-core soft-shoulder (HCSS) potential studied
extensively previously in 2- and 3-D~\cite{r41,r57, r38, r39, r42}, 2) Hard-core
long-range repulsion potential (HCLR), which, to the best of our
knowledge, has not been studied yet in this specific form, but similar
potentials were considered in the past~\cite{r40,r53}, 3) short-range
attraction and long-range repulsion potential (SALR), which was also
studied extensively but in a different form~\cite{r53,r2,r58} , 4) local
attraction soft-shoulder (LASS) potential, which has a minimum but no long-range tail and 5) local
attractive minimum and long-range repulsion (LALR) potential. The last two potentials are considered
here for the first time. The summary of all the employed parameters is shown in table~\ref{t1}.

All systems were studied using the standard Metropolis Monte Carlo (MC) method~\cite{r59}. For a quick scan in
the phase space, simulation boxes with 216 particles were considered at 6 values of the total density.
The densities ranged from 0.17 to 1.69 in reduced units. A number of temperatures were investigated,
depending on the needs of each system, as discussed in the main text. Temperature is reported in units
of $\epsilon_b/k_\text{B}$, where $k_\text{B}$ is the Boltzmann constant. In potentials
with tails, HCLR, LALR and SALR, the
truncation distance $r_\text{c}=2 \sigma_A$
was used. Tests were performed with longer $r_\text{c}$ to make sure that the
results are not affected by this parameter. Additionally, simulations with larger boxes containing 5832
particles were conducted for select thermodynamic points in order to extract cluster statistics. The maximum
displacement of particles in the MC moves was adjusted to achieve 30\% success rate.
All simulations were run for more than 1$\times$10$^6$ MC steps and tests were performed to make sure
that the reported results are converged.

\begin{table}[!h]
\centering
\caption{Parameters of five model
potentials studied in this work. All
energies are measured in units of $\epsilon_b$
and all distances in units of $\sigma_A$.
Abbreviations are as in the main text.
}\label{t1}
\vspace{2ex}
\begin{tabular}{cccc}
\hline\hline
Model & $\epsilon_w$ & $\delta \epsilon$ & $\kappa$ \\
\hline
HCSS & 1 & 0 &  \\
HCLR & 1 & 0 & 4.05 \\
LASS & 0.525 & 0.475 &  \\
LALR & 0.525 & 0.475 & 4.05 \\
SALR & --0.475 & 0.475 & 4.05 \\
\hline\hline
\end{tabular}
\end{table}

To quantify the process of self-association, all the recorded conformations were clustered. The standard
clustering algorithm was used which assigns a particle to the given cluster if its separation from any
particle in the cluster is less than a cut-off distance $R_\text{c}$. For the cut-off, the
diameter of the soft-shoulder $\sigma_A$
was used.

In order to characterize the progress of clusterization, a ratio $x$ of the number of
particles in monomeric
state to the total number of particles was analyzed.
Configurations saved in simulations were used to construct a distribution
$P(x)$ as a function
of density and other thermodynamic parameters. The distribution function yields the free energy profile
$\Delta F(x)=-k_\text{B} T \log{P(x)}$,
where $k_\text{B}$ is the Boltzmann constant and
$T$
is the temperature, which has a
minimum at the most likely value of $x$.
Function $\Delta F(x)$ measures the free energy difference between
conformations with two different values of parameter $x$, for instance 0 and 1, in which case it reports
the full cost/benefit of converting the system at the given thermodynamic point into clusters. The free
energy can be decomposed into internal energy and entropy contributions using the standard thermodynamic
relationship $\Delta F(x) = \Delta U(x) - T \Delta S(x)$.
The unknown functions $\Delta U(x)$ and $-T \Delta S(x)$ can be
determined if the temperature dependence of $\Delta F(x)$ is available. First, numerical differentiation of
$\Delta F(x)$ can yield $-T \Delta S(x)$ (ignoring the
entropy dependence on temperature in the numerical algorithm) and then entropy and
free energy can be combined to produce internal energy. This method is commonly used in literature to
monitor the progress of various biochemical reactions (see for instance reference~\cite{r55} and references therein).
In the analysis of the free energy profiles, the small size of the simulation box was not seen to affect
the conclusions of the work.

\section{Results}
\subsection{All of the studied models form equilibrium clusters}

Several models were studied by computer simulations, as described in detail in the `Methods' section.
\begin{figure}[htbp]
\centerline{
\includegraphics[width=0.51\textwidth]{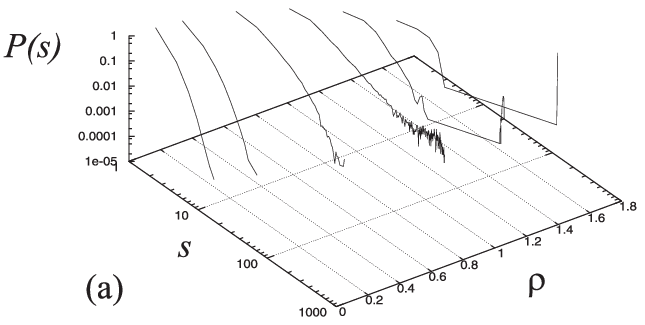}
\includegraphics[width=0.49\textwidth]{clust2.eps}
}
\caption{(Color online)  The number of particles observed in clusters of specific type relative to
the total number of particles in the
simulation box. In (a) data for LALR model are shown for select densities. In (b)
data for all models are shown, each
model at a specific density, indicated between parentheses. All data were collected at
a reduced temperature $T=0.62$.
}
\label{f1}
\end{figure}
All of the studied systems were seen to form equilibrium clusters at appropriate densities and temperature. As
an illustration, figure~\ref{f1}~(a) shows
$P(s)$,
the ratio of particles engaged in clusters of size $s$ to the total
number of particles in the simulation box observed for the HCSS model at a reduced temperature
$T=0.62$. At low density
$\rho=0.17$ (shown in units of
$\sigma_A^{-3}$ ),
the majority of all particles are monomers.
As the density is increased, the proportion of monomers drops at the benefit of dimers, trimers and
other multimers. At $0.34<\rho<0.6$, the population of monomers drops below 0.5, at which point the
system becomes dominated by clusters. Clusterization continues for higher densities, and at
$\rho=1.01$,
the number of monomers drops below the number of particles involved in dimers. At still higher
densities,
$\rho>1.35$, the system undergoes a percolation transition, above which only a single cluster
encompassing the entire simulation box survives. Figure~\ref{f1}~(b) shows
$P(s)$
recorded for different systems
at the density just below the percolation threshold. While all curves demonstrate a majority cluster
population, two specific features stand out. First, the SALR simulations show a strong preference for
cluster size of $\sim$10 monomers. All other systems populate a wide spectrum of sizes with only a small
population maximum for dimers and trimers. Second, in the limit of large $s$, the data of HCSS simulations
demonstrate a clear power-law dependence,
$P(s)\sim 1/s^n$.
The observed exponent of
$n=1.2$
is in
excellent agreement with the prediction of the random percolation
limit~\cite{r39} [note that the usual cluster-size distribution function
$n(s)$ is related to the distribution function used here
as $n(s)=P(s)/s$]. In all other systems, $P(s)$ decays faster.

\subsection{Cluster statistics critically depends on the type of the potential}

\begin{figure}[!t]
\centering
\includegraphics[width=0.5\textwidth]{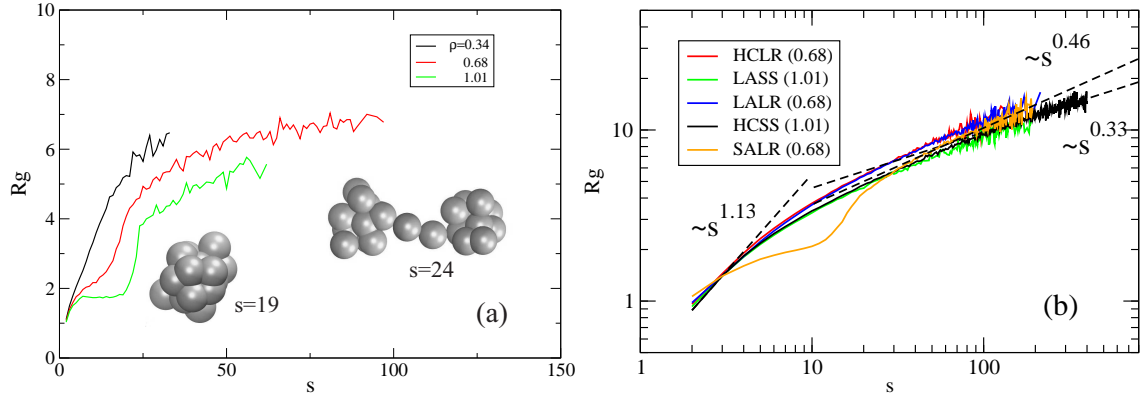}
\includegraphics[width=0.49\textwidth]{rg-a.eps}
\caption{(Color online) Radius of gyration $R_\text{g}$
as a function of the number of particles $s$. Data for SALR potential are shown in
(a). All potentials with appropriate densities are shown in (b).
}
\label{f2}
\end{figure}

All the studied systems populate the clusters that differ widely in size and shape. Radius of gyration
$R_\text{g}(s)$
is
used to characterize how the size of a cluster depends on the number of particles it contains. The most
interesting dependence is observed for the SALR potential. Figure~\ref{f2} shows
$R_\text{g}(s)$
for this model recorded
for several values of the total density of the solution $\rho$.
At low densities, the radius of gyration is a
monotonous function of the number of particles. Starting at $\rho=0.68$,
it begins to develop a plateau,
signaling the onset of a configurational change. At $\rho=1.01$,
	$R_\text{g}$ remains unchanged for $3<s<19$
and then jumps twofold for
$s>23$. It is easy to figure out what happens at the transition point by
examining the shape of the clusters. Representative clusters for
$s=19$,
which is just before the
transition, and
$s=24$,
which is immediately after it, are shown in figure~\ref{f2}~(a). The smaller, or primary,
clusters form dense and almost spherical clumps of particles. The larger clusters are made of two (or
more) such clumps joined together. Thus, cluster formation is driven by hierarchical supramolecular
assembly: small clusters are assembled at initial stages and later on serve as building blocks for larger clusters. This scenario explains the sudden increase in
$R_\text{g}(s)$
at a certain transition value $s_\text{c}$.

The unusual cluster statistics can be rationalized from the standpoint of equilibrium thermodynamics.
In common with other SALR potentials~\cite{r52,r53}, the ground-state (GS) configuration of the current model
is a 1D Bernal spiral. The energy of the spiral is a monotonously decreasing function of $s$
that continues
to decline until all available monomers are absorbed onto the cluster. At zero and sufficiently low
temperature, the spiral is the only observable conformation. At finite temperatures, however, this
picture changes dramatically. The entropy that is lost by the monomers during the assembly process becomes
important. The lowest free energy is achieved through a combination of both internal energy and entropy. The
balance between energy and entropy contributions dictates the size of the primary cluster $s_\text{c}$.
The loss
of energy due to the fracturing of the ground state is compensated for by the gain in the entropy arising
from the translational freedom of the resulting clusters. As the density of the solution increases (volume
is lowered), the entropic gain decreases. So will do the associated loss of energy. As a consequence, the
GS conformation will break into a smaller number of pieces. This means that the size of the primary
cluster should go up with density. This is exactly what one sees in
figure~\ref{f2}~(a) after comparing
$R_\text{g}(s)$ for
$\rho=0.68$ and $1.01$.

The statistics of all other systems are similar to one another and can be described as those of a polydisperse mixture of
clusters of different sizes and shapes. As in the case of SALR, the size of the clusters also decreases with
the increase of density but this effect is much less pronounced. No non-trivial behavior is seen. A
detailed comparison of $R_\text{g}(s)$
obtained for different systems [considered at the same densities as in
figure~\ref{f1}~(b)] is presented in figure~\ref{f2}~(b). The plateau in the SALR curve is seen again.
Moreover, it is
possible to discern the scaling statistics for large clusters. In that limit, the radius of
gyration has a power-law
dependence
$R_\text{g}(s)\sim s^n$
with
$n=0.46$, which is independent of the density. This exponent is close to
$n=0.5$ observed for ideal, or Gaussian, polymer chains which lack the excluded volume interactions. The
Gaussian model mechanism may be applied to the assembly of large colloidal clusters. Although the primary
clusters do have the excluded volume, they are allowed to inter-penetrate via the exchange of particles. Thus,
from the statistical point of view, they act as zero-size particles. For the same reason,
SALR clusters do
not experience compression from the environment, as do HCSS clusters, and remain swollen compared to
maximally compact states.

The statistics of other systems in the large-cluster limit are governed by
$n=0.33$, which corresponds to a
maximally compact object. This observation is illustrated in figure~\ref{f2}~(b)
for HCSS potential. For the same
system in the limit of small $s$, the radius of gyration scales as
$R_\text{g}\sim s^m$,
 where $m=1.13$.
Note that for the
linear arrangement of the smallest clusters, dimer and trimer, the exponent
$m=1.2$. Therefore, HCSS
behaves as a linear chain for small $s$. This is in agreement with the prior work on this
system~\cite{r39}, which also
shows that the linear scaling persists to large clusters for a specific choice of potential parameters
(different from those of the present work). HCLR, LASS and LALR potentials also exhibit almost a linear
scaling but with the exponent that is system-specific. In common with the HCSS model, clusters formed
by these potentials experience a cross-over from a linear regime for small clusters to a collapsed chain
for large clusters. The transition is gradual and takes place over two orders of magnitude. This behavior
is in sharp contrast to the SALR model and can be explained by the specific character of the HCSS interaction.
Although the formation of clusters in this model is driven by entropy (see below), the specific shape of
different clusters of the same size is also influenced by energy. Since the interaction is purely
repulsive, particles within a cluster will choose to reside as far from their neighbors as possible. This
results in linear conformations for small clusters, which minimize the overlap between constituent
particles. For larger clusters, the minimal overlap can be achieved in string-like but bended
conformations, similarly to those observed in the self-avoiding polymers. Like in the polymers, colloidal
clusters should experience an entropic collapse with the pertinent exponent of the radius of gyration of
$m=0.591$, or approximately 3/5 as predicted by Flory~\cite{r54}. The fact that the observed
exponent 0.33 is
lower indicates that large clusters experience additional compression from the environment, which
may only result from the repulsive interactions of clusters with other species in the solution, both
clusters and monomers alike. These theoretical arguments lead us to the conclusion that the specific
details of the cross-over will depend on a) the extent of the soft shoulder in the potential, which governs
how much linear chains gain in energy compared to the collapsed ones and b) the strength of the
repulsive tail, which determines how strongly various colloidal species repel each other, thus inducing cluster collapse.

\section{Discussion}
\subsection{
Two classes of cluster-forming systems use different assembly mechanisms
}
\begin{figure}[!t]
\centering
\includegraphics[width=0.7\textwidth]{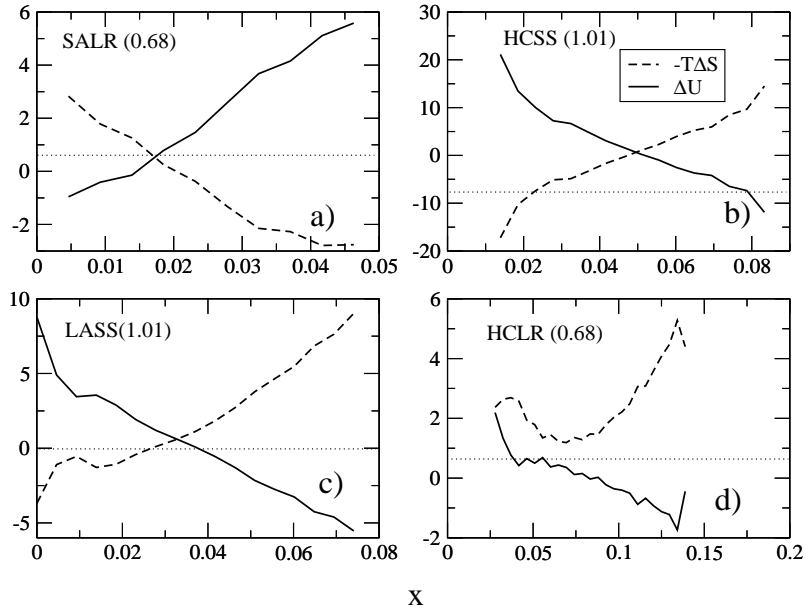}
\caption{
Internal energy, $\Delta U(x)$, and entropy, $-T \Delta S(x)$, as a function of the fraction of
monomeric particles $x$ generated in
this work for different cluster-forming potentials. All models except SALR exhibit
the clusters stabilized by entropy. Here,
and elsewhere in the article, energy is measured in units of $\epsilon_b$.
}
\label{f3}
\end{figure}

Our simulations clearly demonstrate that all of the studied systems form equilibrium clusters.
In view of this
finding, the relevant question to ask is: `Why does that happen? What are the pertinent mechanisms?'.
Perhaps the simplest to explain is the mechanism of the SALR model. By design, this potential favors
association of particles by providing a negative potential energy to configurations with close separation
among particles. Clusters of different topologies represent a ground-state configuration of SALR
potentials~\cite{r2,r52}, including the one studied here. Thus, for these systems,
cluster assembly is expected to
be driven by internal energy. To quantify this assessment, it makes sense to analyze the free energy profile of the
clusterization reaction as a function of a certain order parameter. Since we are interested in the
transition into the cluster state in general, not into a specific type of cluster, it is most convenient to
monitor the number of particles in the monomeric state as a progress variable. To make a comparison
between systems of different sizes easier, the number of monomers will be normalized by the total
number of particles, producing a ratio of monomers $0<x<1$.
Conformations with no clusters
have $x=1$
 while those lacking monomers (complete transformation into clusters) are characterized by
$x=0$.
Internal energy,  $\Delta U(x)$, and entropy, $-T \Delta S(x)$, can be computed for each $x$ (see `Methods'
section)
to elucidate the role of these functions in the cluster formation.
Figure~\ref{f3}~(a) shows $\Delta U(x)$ and $-T \Delta S(x)$ obtained for the SALR model at
$\rho=0.68$. As predicted above,
the cluster formation is driven by energy and opposed by entropy. The two curves combined yield a
minimum of free energy at about
$x=0.015$.

\begin{figure}[!b]
\centering
\includegraphics[width=0.35\textwidth]{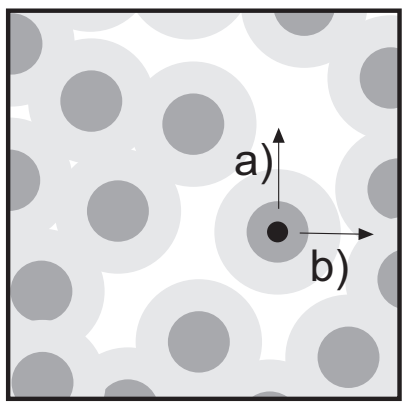}
\caption{
Illustration of how entropy stabilizes the clusters in the HCSS model. The central monomeric particle,
denoted by black circle, has the choice of either to: a) remain monomeric, in which case it is allowed to
explore white-colored space, or b) join a cluster, in which case it is allowed to explore light-gray space.
Dark gray space indicates hard-core areas which are not accessible
to the particle. Clusters become stable when
the volume of the light-gray area exceeds that of the white area.
}
\label{f4}
\end{figure}

The same analysis was repeated for the HCSS potential. Note that this model in 2D, given the
appropriate choice of parameters, forms ordered configurations in the ground and low-energy states
\cite{r56} as well as quasi-crystals~\cite{r46} .
Thus, one may expect the remnants of ordered structures to survive at finite
temperature, giving rise to clusters stabilized by energy~\cite{r42}. Figure~\ref{f3}~(b),
depicting $\Delta U(x)$ and
$-T \Delta S(x)$ demonstrates that this expectation is not justified. Clusters are
stabilized entirely by entropy
and destabilized by energy. The two systems, SALR and HCSS, differ dramatically in their cluster
formation scenarios. While clusters in SALR tend to stay intact due to the mutual attraction, clusters in
the HCSS model use a different mechanism. Namely, particles at distances
$\sigma_\text{H} < r < \sigma_A$
experience
no force and move freely about each other. Particles in the monomeric state experience no force
either, since the potential is zero for $r > \sigma_A$.
Therefore, clusters stay together because the volume created by
joining the soft-shoulder regions of all constituent particles is greater than the volume available to them
in the monomeric state (at a distance $r > \sigma_A$
from any other monomer or cluster). This point is
illustrated in figure~\ref{f4}. For a monomeric particle, the moment of joining an existing cluster is accompanied by an increase
in energy as well as by an increase in entropy. As long as the entropy wins, particles spend the majority of
their time in cluster configurations. Note that these clusters will have a very short residence time of the
order of the duration of binary collisions. They will also rapidly disintegrate in the event of an increase in the available volume.

Cluster formation in HCRL and LASS models is driven by the same mechanism as in the HCSS model.
Adding a repulsive tail to the potential, as in HCLR, changes the appearance of $-T \Delta S(x)$: at
small $x$ it
begins to disfavor clusters, see figure~\ref{f3}~(d). However, the destabilizing effect
of energy is unaffected. An
additional minimum in the soft shoulder, as in LASS, decreases the destabilizing effect of energy,
compare figure~\ref{f3}~(c) and (b), but not to the extent of reversing its role.
Cluster formation of both systems
is driven by entropy.

\subsection{The model with local minimum exhibits a dual identity}

\begin{figure}[!h]
\centering
\includegraphics[width=0.55\textwidth]{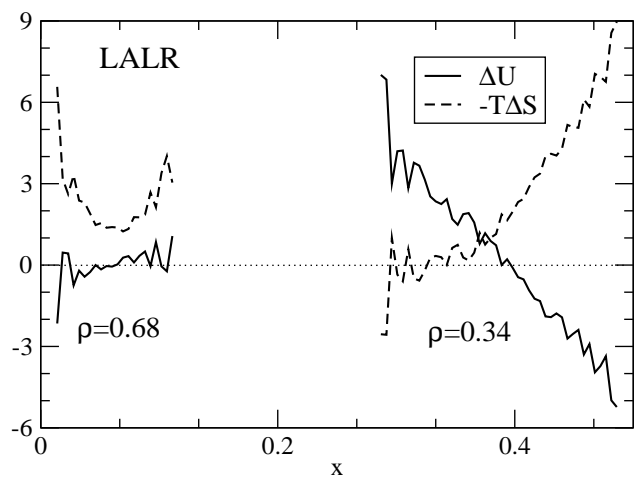}
\caption{
The same as figure~\ref{f3}, but for the potential with local minimum and repulsive tail, LALR. Both entropy- and
energy driven mechanisms of cluster formation are present, depending on the total density of the solution.
}
\label{f5}
\end{figure}

The model with the local minimum and repulsive tail stands out from the rest of the studied systems.
Figure~\ref{f5} shows its energy and entropy evaluated at two different densities,
$\rho=0.34$ and
$\rho=0.68$. At a lower density, the mechanism is as in the HCSS system with the entropy stabilizing the clusters. For
a higher density, the mechanism changes and becomes similar to that of the SALR potential, where the internal energy drives the assembly of clusters. The transformation is not complete as entropy still favors
the clusters at high density instead of disfavoring them, as in the SALR potential. Nevertheless, the reversal
of the energy role is quite remarkable. This is not seen in any other potential and indicates that the
process of cluster formation in the systems interacting via potentials with local minima can be
extraordinarily complex. It takes the presence of both monomers and small clusters in such systems
 for the clusterization to be driven by energy. Any theory that aims at predicting the  cluster distribution
quantitatively should be capable of capturing that effect. That means a proper description of
monomer/cluster mixture, including the association/dissociation balance, across a wide range of
densities. It is not clear which of the currently available liquid-state theories would be capable of accomplishing this task.

\begin{figure}[!t]
\centering
\includegraphics[width=0.5\textwidth]{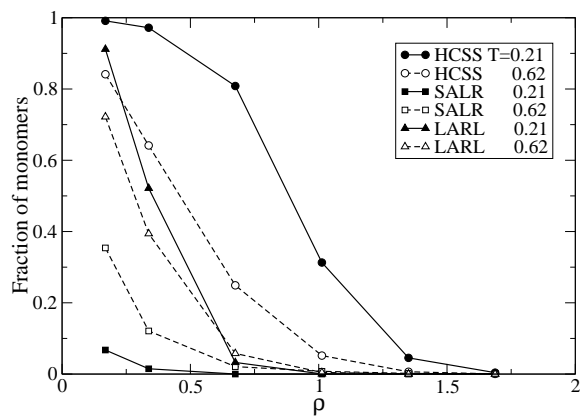}
\caption{
Fraction of monomers in the conformational ensemble obtained for different models, temperatures
and densities. HCSS clusters are destabilized by temperature. SALR clusters are stabilized by temperature
while LALR clusters exhibit both behaviors depending on density.
}
\label{f6}
\end{figure}

The split identity of the LALR model is also evident in its temperature behavior. Since the assembly of
clusters in the HCSS potential is entropic in nature, it should be enhanced by temperature. The opposite
is true for the SALR model, where clusters should become less stable at higher temperature.
Figure~\ref{f6},
showing the fraction of monomers as a function of temperature and density for different models,
confirms this conjecture. Different symbols in this figure represent different systems, broken lines
correspond to a high temperature of 0.62 while solid lines stand for the low temperature of 0.21. For
the HCSS model, the population of monomers at the lower temperature is higher than the population at
a higher temperature, the solid line is above the broken line, at all densities, indicating that cluster
formation is enhanced by temperature. For the SALR model, the broken line is above the solid line, which is
the evidence that temperature hinders the formation of clusters. For the LALR potential, the broken line is
above the solid line for
$\rho<0.68$ and vice versa for higher densities. Thus, this model exhibits the features
characteristic of both classes of cluster-forming colloids. The entropy-driven mechanism dominates at
low densities while the energy-driven mechanism~--- at high densities.

\section{Conclusions}
There are two known types of cluster-forming colloids: a) those that
interact via purely repulsive
potential with a soft shoulder and, possibly, a tail, and b) those that are
characterized by
a potential with the global attractive well at short distances followed by a repulsive tail at
long distances.
By adding an attractive term to the otherwise repulsive potential, it is easy to transform the potentials
of the first type into the potentials of the second type. Thus, it may seem obvious
that systems of the two types exhibit similar properties, in particular the capability of forming clusters.
However,
there is one caveat with this argument.
Extrapolating properties of one system based on the results of another system makes sense only
if the two are sufficiently similar so that perturbation theory can be used. This is clearly not
the case for the cluster-forming colloids. Repulsive and attractive potentials exhibit completely
different structural, dynamic and phase behaviors. Take, for instance, the gas-liquid transition in
attractive colloids that is missing in the repulsive colloids. The differences also extend to the
clusters of these systems. While attractive potentials form compact clusters of specific size, the
repulsive potentials lead to extended, polydisperse multimers. There is only a limited amount of information that
 can be learned about one system by extrapolating the properties of the other.
Simulations remain the most reliable and accurate tool in the studies of a new or unknown potential.
Simulations played a key role in this work, helping us
 to reveal that a repulsive potential with a local minimum shares
certain properties with both other types of cluster-forming potentials.
The clusters it populates are similar to those of purely repulsive potentials, while the energy-driven
mechanism is similar to the attractive potentials.

Our results were obtained for a specific shape of the potential with a local minimum. How
general they are with respect to other systems remains to be seen.
Of particular interest here would be to consider
square-well potentials which have been studied extensively
in the literature in regard to phase transitions.
Also, of interest would  be to examine the behavior of continuous potentials, like
those that can be constructed with the help of two Yukawa functions.


\ukrainianpart
\title{Рівноважні кластери в розчинах колоїдів, що взаємодіють через потенціал
з локальним мінімумом}

\author{А. Баумкетнер\refaddr{label1}, В. Кай\refaddr{label2}}
\addresses{
\addr{label1} Інститут фізики конденсованих систем НАН України, вул. І.~Свєнціцького, 1, 79011 Львів, Україна
\addr{label2} Дослідний центр обчислювальних наук у м. Пекін, 1000094 Пекін, Китай
}

\makeukrtitle

\begin{abstract}
В простих колоїдних розчинах кластерами називаються мультимери, що виникають
внаслідок асоціації мономерів і співіснують з ними в динамічній рівновазі.
Існують два типи потенціалів, що ведуть до утворення кластерів: а) потенціали з
глобальним мінімумом, що можуть
виникнути, зокрема, внаслідок неповної компенсації притягальної взаємодії на коротких
відстанях та відштовхувальної взаємодії на довгих відстанях, б) повністю відштовхувальні
взаємодії, які демонструють м'яке ``плече''. За допомогою комп'ютерних симуляцій
в даній роботі показано, що потенціали з локальним мінімумом і відштовхувальним
хвостом теж мають здатність
утворювати рівноважні кластери. Хоча вони не належать до жодного з перелічених
вище класів, аналіз показує, що такі
потенціали мають з ними певні спільні характеристики. При малій густині колоїдів
кластери, утворені  новими
потенціалами,
видовжені подібно до кластерів в суто відштовхувальних системах. При великій густині, нові
кластери компактні, як це спостерігається в системах з глобальним потенціальним мінімумом.
Отже, нові потенціали слугують тим містком, який поєднує два відомі класи потенціалів, що
ведуть до утворення кластерів в колоїдних розчинах.

\keywords колоїди, кластери, локальний мінімум, відштовхувальний потенціал, \\ комп'ютерні
симуляції

\end{abstract}

\lastpage

\end{document}